\documentclass[
twocolumn,
floatfix,
prb,
longbibliography,
superscriptaddress
]{revtex4-2}
\usepackage[dvips]{graphicx}
\usepackage{bbm}
\usepackage{amsmath}
\usepackage{amsfonts}
\usepackage{lipsum}
\usepackage[colorlinks=true,allcolors=blue]{hyperref}
\usepackage{ulem}

\allowdisplaybreaks

\begin{document}

\title{Majorana-mediated thermoelectric transport in multiterminal junctions}

\author{Raffael~L.~Klees}
\thanks{These authors contributed equally to this work.}
\affiliation{Institute for Theoretical Physics and Astrophysics and W{\"u}rzburg-Dresden Cluster of Excellence ct.qmat,\\
Julius-Maximilians-Universit{\"a}t W{\"u}rzburg, D-97074 W{\"u}rzburg, Germany}

\author{Daniel~Gresta}
\thanks{These authors contributed equally to this work.}
\affiliation{Institute for Theoretical Physics and Astrophysics and W{\"u}rzburg-Dresden Cluster of Excellence ct.qmat,\\
Julius-Maximilians-Universit{\"a}t W{\"u}rzburg, D-97074 W{\"u}rzburg, Germany}

\author{Jonathan~Sturm}
\affiliation{Institute for Theoretical Physics and Astrophysics and W{\"u}rzburg-Dresden Cluster of Excellence ct.qmat,\\
Julius-Maximilians-Universit{\"a}t W{\"u}rzburg, D-97074 W{\"u}rzburg, Germany}

\author{Laurens W. Molenkamp}
\affiliation{Experimental Physics III, Julius-Maximilians-Universit{\"a}t W{\"u}rzburg, D-97074 W{\"u}rzburg, Germany}
\affiliation{Institute for Topological Insulators, Julius-Maximilians-Universit{\"a}t W{\"u}rzburg, D-97074 W{\"u}rzburg, Germany}

\author{Ewelina~M.~Hankiewicz}
\affiliation{Institute for Theoretical Physics and Astrophysics and W{\"u}rzburg-Dresden Cluster of Excellence ct.qmat,\\
Julius-Maximilians-Universit{\"a}t W{\"u}rzburg, D-97074 W{\"u}rzburg, Germany}

\begin{abstract}
The unambiguous identification of Majorana zero modes (MZMs) is one of the most outstanding problems of condensed matter physics.
Thermal transport provides a detection tool that is sensitive to these chargeless quasiparticles. 
We study thermoelectric transport between metallic leads transverse to a Josephson junction.
The central double quantum dot hosts conventional or topological Andreev states that depend on the phase difference $\phi$.
We show that the presence of MZMs can be identified by a significant amplification of both the electrical and thermal conductance at $\phi \approx \pi$ as well as the Seebeck coefficient at $\phi \approx 0$.
In addition, we show that the Wiedemann-Franz law is strongly violated in the presence of MZMs around $\phi \approx \pi$ when compared to the conventional case.
We further investigate the robustness of our results against Cooper pair splitting processes.
\end{abstract}

\date{\today}
\maketitle
\section{Introduction}
Josephson junctions (JJs) have been extensively studied in numerous works, driven by their wide range of applications, from metrology \cite{Fatemi2021,belcher2018} and quantum simulation \cite{manousakis2002quantum} to quantum computation \cite{Flensberg2011,stern2013topological,sarma2015majorana}.
Recently, topological JJs gained significant attention as they provide robust platforms hosting Majorana zero modes (MZMs) 
\cite{Fernando2014Robustsignatures,sato2017topological,Ren2019,Fornieri2019}. 
In particular, similar quantum-dot-based setups have demonstrated promising potential as platforms for flying qubits \cite{Waintal2018,Zhang2018}, as these systems feature well-established quantum interferometers 
\cite{Heiblum1997,Cleuziou2006,veldhorst2012experimental}.

Even though the unequivocal detection of MZMs remains an open problem, several approaches have been proposed to investigate the topological nature of JJs, such as analyzing the current-phase relation \cite{Rodriguez2022,Balseiro_PRB_19,vecino2003,platero,Grun2019,ren2018measuring}.
While the quantized electrical conductance was initially considered an exclusive feature of topological materials, it was later realized that it can arise from any zero-energy mode \cite{alicea2012new,akhmerov2011quantized,yu2021non,sengupta2001midgap}.

Since the establishment of standard thermoelectric measurement techniques in the early 1990's \cite{Molenkamp1992,vanHouten1992}, they have advanced to promising tools to detect chargeless MZMs, with their signatures manifested in the thermal conductance \cite{fu2008superconducting,Bauer2021}, voltage thermopower \cite{dolgirev2019topology,Hou2013ThermoMottMajo,Sela2019FractionalEntropy}, or the violation of the Wiedemann-Franz law \cite{Giuliano2022,Buccheri2022,Ritesh2022}.
In quantum-dot-based multiterminal setups, it was reported that a possible smoking gun to detect MZMs is an opposite sign-behavior of the Seebeck coefficient as a function of the energy level of the quantum dot (QD) compared to a conventional superconductor \cite{Lopez2014,Weymann2017,RamosAndrade2016,Ricco2018,Sun2021,Valentini2015}.

\begin{figure}
	\centering
	\includegraphics{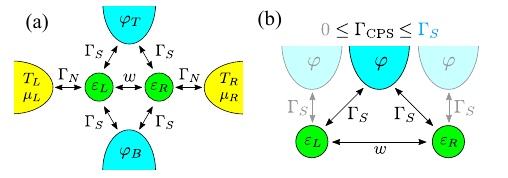}
	\caption{
		(a) 
		Two QDs (green) with energy levels $\varepsilon_{L,R}$ and interdot coupling $w$.
		The left (L) [right (R)] normal-metal terminal (yellow) at temperature $T_L$ [$T_R$] and chemical potential $\mu_L$ [$\mu_R$] is coupled to the left [right] QD with coupling $\Gamma_N$.
		The top (T) and bottom (B) superconductors (cyan) with phases $\varphi_{T,B}$ are equally coupled to both dots with strengths $\Gamma_{S}$.
		(b) CPS is controlled with the nonlocal coupling $\Gamma_{\mathrm{CPS}}$.
		If $\Gamma_{\mathrm{CPS}} = 0$, each QD is coupled to a copy of the superconductor with identical phase $\varphi$.
	}
	\label{fig:figure1}
\end{figure}

In this paper, we study the multiterminal system depicted in Fig.~\ref{fig:figure1}(a), which also represents a generalization Cooper pair splitting (CPS) setups \cite{Hofstetter2009,Herrmann2010,Schindele2012,Cao2015}.
The central region consists of two coupled QDs, each of them connected to one metallic lead and both are connected to two superconducting leads.
We restrict our analysis to the noninteracting limit since it has already been shown that the low-energy behavior and the conductance is dominated by the effects of MZMs even in the Kondo regime \cite{Majek2022,Lara2023}.
In contrast to previous discussions aforementioned, we find that the presence of a MZM does not generally result in the sign change of the Seebeck coefficient.
Therefore, alternative methods are required for their definitive identification.
For this purpose, we propose a measurement of the transverse thermoelectric transport coefficients as a function of the phase difference $\phi = \varphi_B - \varphi_T$ between the bottom (B) and top (T) superconductors.
In particular, we show that the linear-response signals of both the electrical and thermal conductance around $\phi \approx \pi$ as well as the Seebeck coefficient around $\phi \approx 0$ show a huge amplification in the presence of MZMs.
We also show that MZMs lead to a strong violation of the Wiedemann-Franz law around phase differences $\phi \approx \pi$.

\section{Double-quantum-dot model}
We consider the four-terminal junction shown in Fig.~\ref{fig:figure1}(a) with a central spin-degenerate, noninteracting double QD.
\begin{figure}
	\centering
	\includegraphics{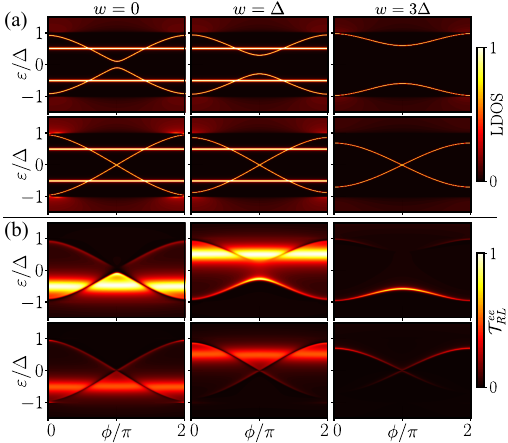}
	\caption{
		(a) LDOS on the double QD coupled to either two $s$-wave (top row) or two $p$-wave (bottom row) superconductors with $\Gamma_N = 0$.
		(b) EC transmission $\mathcal{T}^{ee}_{RL}$ with $\Gamma_N = \Delta / 5$ for two $s$-wave (top row) or $p$-wave (bottom row) superconductors.
		Parameters: $\Gamma_{S} = \Delta$, $\Gamma_{\mathrm{CPS}} = \Gamma_{S}$, $\varepsilon_L = \varepsilon_R = -\Delta / 2$.
        }
	\label{fig:figure2}
\end{figure}
The central Hamiltonian reads $H_{\mathrm{QD}} = \sum_{\alpha,\sigma} \varepsilon_\alpha d_{\alpha\sigma}^\dagger d_{\alpha\sigma}$, where $d_{\alpha\sigma}^{(\dagger)}$ annihilates (creates) an electron of spin $\sigma \in \{ \uparrow, \downarrow \}$ at the onsite energy $\varepsilon_\alpha \in \mathbb{R}$ of QD $\alpha \in \{L,R\}$ and the interdot coupling of strength $w \geq 0$ is given by $V_{\mathrm{QD}} = -w \sum_\sigma (d_{L\sigma}^\dagger d_{R\sigma} + d_{R\sigma}^\dagger d_{L\sigma} )$.
The full model including the leads is represented by the Hamiltonian $H = H_\mathrm{QD} + V_{\mathrm{QD}} + H_\mathrm{leads} + H_\mathrm{tunnel}$, where $H_\mathrm{leads}$ describes the four terminals modeled by semi-infinite chains \cite{cuevas2nd,Yeyati_PRB_16}, with details provided in Appendixes~\ref{sec:appendixA} and \ref{sec:appendixB}.
The the dot-to-leads coupling is described by $H_\mathrm{tunnel}$, characterized by the parameter $\Gamma_{N,S}$ for normal and superconducting leads, respectively.
To study the effect of topological superconductivity on transverse thermoelectric transport, the superconductors are either of conventional $s$-wave or topological $p$-wave type.
For simplicity, the latter case is modeled by semi-infinite spinless Kitaev chains in the deep topological regime \cite{Yeyati_PRB_16,kitaevchain}, which are members of symmetry class BDI \cite{Bespalov2023}. 
More realistic models based of a semiconducting nanowires with strong spin-orbit coupling in a magnetic field, which belong to symmetry class D \cite{Chiu2016}, should yield the same results in the fully spin-polarized (i.e., effectively spinless) regime, in which Majorana end states appear.

As sketched in Fig.~\ref{fig:figure1}(b), the parameter $\Gamma_{\mathrm{CPS}}$ describes the nonlocal splitting of Cooper pairs into two electrons, each being transferred to a different QD.
In general, $\Gamma_{\mathrm{CPS}}$ depends on the geometry of the contact and the coherence length of the Cooper pairs \cite{Recher2001} and allows the modeling of more realistic situations. 

\section{Transmission functions and band inversion}
We study the transmission functions between the two normal-metal electrodes as an effective two-terminal setup.
Due to the connection to the superconductors, the center hosts Andreev bound states (ABS) that mediate the transverse thermoelectric transport. 
As shown in Appendix~\ref{sec:appendixC}, the transmission functions between the terminals $\ell_1,\ell_2 \in \{ L,R \}$ in the spinless $p$-wave case read
\begin{align}
    \label{eq:equation1}
    \mathcal{T}_{\ell_1 \ell_2}^{\tau_1 \tau_2}(\varepsilon) = 4 \, \Gamma_{N}^2 \,  G^r_{\ell_1\tau_1, \ell_2  \tau_2} (\varepsilon) \, G^a_{\ell_2  \tau_2 , \ell_1 \tau_1 }(\varepsilon) ,
\end{align}
where $G^{r,a}$ is the dressed retarded/advanced Green's function of the center and $\tau_1, \tau_2 \in \{e,h\}$ are the electron-hole components.
For two $s$-wave terminals, all Green's functions are spin-symmetric due to the absence of spin-flip terms, which results in an additional factor of 2 in the transmission functions in Eq.~\eqref{eq:equation1}.
Due to particle-hole symmetry, the transmission functions satisfy $\mathcal{T}_{\ell_1 \ell_2}^{\tau_1 \tau_2}(\varepsilon) = \mathcal{T}_{\ell_1 \ell_2}^{\bar{\tau}_1 \bar{\tau}_2}(-\varepsilon)$ with $\bar{\tau} = h,e$ if $\tau = e,h$.
This allows us to focus on electron cotunneling (EC) $\mathcal{T}_{RL}^{ee}$, local Andreev reflection (LAR) $\mathcal{T}_{RR}^{eh}$, and crossed Andreev reflection (CAR) $\mathcal{T}_{RL}^{eh}$, which determine the transverse thermoelectric properties of the junction [cf.~Eqs.~\eqref{eq:equation2}-\eqref{eq:equation4} below].

The main differences between conventional $s$-wave and topological $p$-wave terminals are best observed in the EC transmission function $\mathcal{T}_{RL}^{ee}$ and the local density of states (LDOS) of the central region defined as $ \mathrm{Tr}[ \mathrm{Im}(G^a)]$, where the trace runs over site and particle-hole degrees of freedom. 

We first show in Fig.~\ref{fig:figure2}(a) the LDOS as a function of the phase difference $\phi$ for strong CPS ($\Gamma_{\mathrm{CPS}} = \Gamma_{S}$) and symmetric dot levels ($\varepsilon_L = \varepsilon_R$).
For both conventional and topological leads, we find a single pair of non-dispersive states and a single pair of ABS at energies $|\varepsilon| < \Delta$, where $\Delta$ is the order parameter of the two superconductors.
While the ABS are changing with $\phi$, the constant states are fixed at energies $\varepsilon \approx \pm (\varepsilon_{\mathrm{avg}} + w)$ due to the strong interference enabled by $\Gamma_{\mathrm{CPS}} = \Gamma_S$, where $\varepsilon_{\mathrm{avg}} = (\varepsilon_L + \varepsilon_R)/2$ is the average energy level of the double QD.
Generally, the ABS in the $s$-wave case are gapped around $\phi = \pi$, while there is always a protected MZM at $\phi = \pi$ for $p$-wave terminals.

In Fig.~\ref{fig:figure2}(b), we show the EC transmission function $\mathcal{T}_{RL}^{ee}$.
In both cases, the electronic constant resonant state shows a strong transmission at $\varepsilon \approx \varepsilon_{\mathrm{avg}} + w$ and increasing $w > 0$ leads to a shift of the constant resonant line toward positive energies.
Pushing these states further to energies $|\varepsilon| > \Delta$ reveals the resonant behavior of the dispersing ABS. 
While there is strong electron transmission for the ABS at negative energy in the $s$-wave case, the $p$-wave configuration shows a dominant electron transmission at positive energies.
This relative inverse behavior is a strong signature of band inversion in topological $p$-wave superconductors. 
However, the overall behavior in both cases is not fixed to be particlelike or holelike and can be changed by changing the QD energies.
In the symmetric case $\varepsilon_L = \varepsilon_R$, the inversion point follows the simple condition $w = |\varepsilon_\mathrm{avg}|$, as shown in Fig.~\ref{fig:figureS1} in Appendix~\ref{sec:appendixD}.

For asymmetric values, $\varepsilon_L \neq \varepsilon_R$, the constant energy states hybridize with the dispersing ABS.
However, as shown in Fig.~\ref{fig:figureS2} in Appendix~\ref{sec:appendixD}, the transmission function is then still dominated by a resonant line at energies $\varepsilon \approx \varepsilon_{\mathrm{avg}} + w$.
This neither changes the qualitative behavior nor our previous discussion.
Finally, smaller $\Gamma_{\mathrm{CPS}} < \Gamma_S$, as being of relevance in CPS experiments \cite{Hofstetter2009,Herrmann2010,Schindele2012}, will lead to the appearance of a second pair of dispersing ABS, while the formerly constant energy states start to merge with the continuum $|\varepsilon| > \Delta$.
In that sense, this parameter interpolates between double-dot and effective (multilevel) single-dot behavior, the latter defined by $\Gamma_{\mathrm{CPS}} = 0$ [cf.~Fig.~\ref{fig:figure1}(b)].

\section{Thermoelectric response}
We are interested in the thermoelectric charge and heat currents, $I_R$ and $J_R$, respectively, in the right contact within linear response at the Fermi energy $\mu = 0$.
We consider the small voltage and temperature bias, $\delta V$ and $\delta T$, respectively, to be applied to the right terminal, which implies $T_L = T$, $\mu_L = 0$, $T_R = T + \delta T$, and $\mu_R = e\, \delta V$.
Then, starting from Eqs.~\eqref{eq:transmissionFunctions} and \eqref{eq:thermoelectricCurrent4} in Appendix~\ref{sec:appendixC}, the linear response Onsager relations read \cite{benenti2017fundamental}
\begin{align}
    \label{eq:equation2}
    \begin{pmatrix}
        I_R / e\\ J_R / (k_B T) 
    \end{pmatrix}
    =
    \frac{1}{h}
    \begin{pmatrix}
        \mathcal{L}_{11} & \mathcal{L}_{12} \\
        \mathcal{L}_{21} & \mathcal{L}_{22} \\
    \end{pmatrix}
    \begin{pmatrix}
        e \,  \delta V \\ k_B \, \delta T
    \end{pmatrix} ,
\end{align}
where $\mathcal{L} = (\mathcal{L}_{mn})_{m,n=1,2}$ is the Onsager matrix with
\begin{align}
	\label{eq:equation3}
    \mathcal{L}_{mn} 
    = 
    \int_{-\infty}^\infty  \left(\frac{\varepsilon}{k_B T}\right)^{m+n-2} \mathcal{T}_{mn}(\varepsilon) 
    \left( - \frac{\partial f}{\partial \varepsilon} \right) d\varepsilon .
\end{align}
Here, $f(\varepsilon) = (1 + e^{\varepsilon / (k_B T)} )^{-1}$ is the equilibrium Fermi function and $T,k_B,e,h > 0$ are the temperature, Boltzmann constant, elementary charge, and Planck constant, respectively.
Using Eq.~\eqref{eq:equation1} and particle-hole symmetry, the total transmission functions read
\begin{subequations}
\label{eq:equation4}
\begin{align}
    \mathcal{T}_{11} &= \mathcal{T}_{21} = 
    \mathcal{T}_{RL}^{ee} 
    +  \mathcal{T}_{RL}^{eh} 
    + 2\mathcal{T}_{RR}^{eh}  ,
    \\
    \mathcal{T}_{12} &= \mathcal{T}_{22} =
    \mathcal{T}_{RL}^{ee} 
    +  \mathcal{T}_{RL}^{eh}  .
\end{align}
\end{subequations}
The linear electrical conductance and the Peltier coefficient, measured at thermal equilibrium $\delta T = 0$, read $G = G_0 \mathcal{L}_{11}$ and $\Pi = k_B T \mathcal{L}_{21} / (e \mathcal{L}_{11})$, respectively, where $G_0 = e^2 / h$ is the electrical conductance quantum. 
The linear thermal conductance and Seebeck coefficient, measured at $I_R = 0$, read $K = 3 K_0 \mathrm{det}(\mathcal{L}) / (\pi^2 \mathcal{L}_{11})$ and $S = k_B \mathcal{L}_{12} / (e \mathcal{L}_{11})$, respectively, where $K_0 = \pi^2 k_B^2 T / (3h)$ is the thermal conductance quantum \cite{benenti2017fundamental}.
Since electrons and holes from the same terminal see the same QD, $\mathcal{T}_{RR}^{eh}$ will always be a symmetric function of energy $\varepsilon$ resulting in $\mathcal{L}_{12} = \mathcal{L}_{21}$ and $\Pi = T S$.

Note that the Wiedemann-Franz law \cite{Franz1853}, $K \propto G$, which would imply $\mathcal{L}_{22} \propto \mathcal{L}_{11}$ if $\mathcal{L}_{12} \ll \mathcal{L}_{11}$ at low temperature, is generally violated due to LAR contributions to $I_R$ if the voltage bias is applied to the right terminal \cite{Ghanbari2011,benenti2017fundamental}.
In contrast, a temperature gradient does not generate LAR contributions to the thermal current $J_R$.

\begin{figure}
	\centering
	\includegraphics{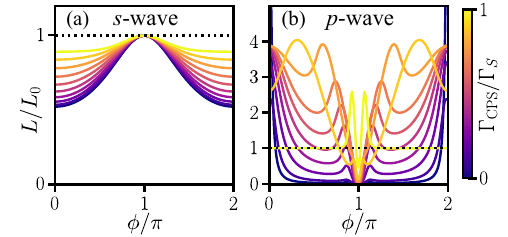}
	\caption{
		Violation of the Wiedemann-Franz law as a function of the phase difference $\phi$ for several values $0 \leq \Gamma_{\mathrm{CPS}} \leq \Gamma_S$. 
		We plot $L := K/(G T) = 3 \, L_0 \, \mathrm{det}(\mathcal{L}) /(\pi^2 \mathcal{L}_{11}^2)$, where $L_0 = \pi^2 k_B^2 / (3e^2)$ is the Lorenz number.
		The Wiedemann-Franz law is satisfied for $L=L_0$ (dotted line).
		Parameters: 
		$w = 3\Delta$,
		$\varepsilon_L = -\Delta/2$, 
		$\varepsilon_R = 0$,
		$k_B T = 10^{-2}\Delta$,
		$\Gamma_N = \Delta/5$,
		$\Gamma_S = \Delta$.
	}
	\label{fig:figureS3}
\end{figure}

For completeness, we show this violation in Fig.~\ref{fig:figureS3} for different phase differences $\phi$ in the $s$- and $p$-wave case.
In the $s$-wave case, shown in Fig.~\ref{fig:figureS3}(a), this violation is always present at phases $\phi \neq \pi$ and becomes largest at $\phi = 0$.
At $\phi = \pi$, the Wiedemann-Franz law is satisfied due to the symmetric choice of all couplings to the superconductors, which results in vanishing CAR and LAR transmission functions. 
Furthermore, the qualitative behavior of the violation does not change between weak or strong CPS controlled by $\Gamma_\mathrm{CPS}$.
In contrast, the $p$-wave case [Fig.~\ref{fig:figureS3}(b)] shows a strong violation for all phases (except some isolated points where $L = L_0$), and in particular for $\phi \approx \pi$, as long as $\Gamma_\mathrm{CPS} < \Gamma_S$.
In the strong CPS case $\Gamma_\mathrm{CPS} = \Gamma_S$, the violation is only appearing around the phase difference $\phi \approx \pi$, i.e., the region close to the MZM.
From this violation it can be concluded that the MZM leads to finite LAR contributions at low energy.

\subsection{Seebeck coefficient}
The Seebeck coefficient $S$ is a measure of whether the thermoelectric transport through a system is particle- or hole-dominated. 
\begin{figure}
    \centering
    \includegraphics{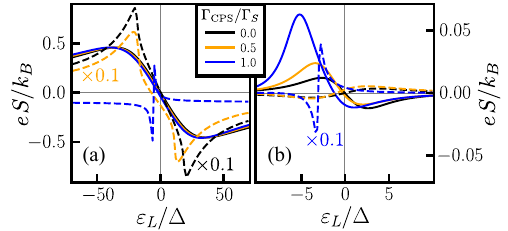}
    \caption{
        Seebeck coefficient $S$ for varying $\varepsilon_L$ at different $\Gamma_{\mathrm{CPS}}$.
        (a) $s$-wave at $\phi = \pi$ (solid) and $p$-wave at $\phi = 0$ (dashed). The orange and black dashed lines are scaled down by a factor 0.1.
        (b) $s$-wave at $\phi = 0$ (solid) and $p$-wave at $\phi = \pi$ (dashed). 
        The blue dashed line is scaled down by a factor 0.1.
        Parameters: $w = 3\Delta$,
		$\Gamma_N = \Delta / 5$, 
		$\Gamma_{S} = \Delta$,
		$\varepsilon_R = 0 \, \Delta$,  
		$k_B T = 10^{-2}\Delta$.
    }
    \label{fig:figure3}
\end{figure}
In Fig.~\ref{fig:figure3}, we first discuss the influence of band inversion on $S$ for both the $s$- and $p$-wave case (solid and dashed curves, respectively).
To understand the effect of the MZM in the p-wave case, we turn our attention to the two characteristic phase differences $\phi = 0$ and $\phi = \pi$, for which $S$ is largest as a function of $\varepsilon_L$.
To distinguish between Majorana-mediated physics and nonlocal effects, we also consider several values of the nonlocal Cooper pair splitting $\Gamma_{\mathrm{CPS}}$.

In the absence of pair-breaking effects ($\Gamma_{\mathrm{CPS}} = 0$), the Seebeck coefficient in the p-wave case [black dashed curve in Fig.~\ref{fig:figure3}(b)] is inverted compared to the corresponding s-wave case [black solid curve in Fig.~\ref{fig:figure3}(a)]; an effect solely caused by the presence of MZMs and band inversion in the topological junction. 

On a qualitative level, increasing $\Gamma_{\mathrm{CPS}}$ does not change too much the general behavior of $S$ in the $s$-wave case, while there is a strong dependence for two $p$-wave terminals.
In the latter case for strong $\Gamma_{\mathrm{CPS}}$ (dashed blue lines in Fig.~\ref{fig:figure3}), the magnitude of the Seebeck coefficient is the same regardless of the phase difference and the region in which the sign change appears is very narrow. 
This robust behavior can be explained with the presence of the resonant non-dispersing level that appears at the energy $\varepsilon \approx \varepsilon_{\mathrm{avg}} + w$. 
The only qualitative difference between the dashed blue lines in Fig.~\ref{fig:figure3} is the presence of the MZM at $\phi = \pi$ [Fig.~\ref{fig:figure3}(b)] with a large conductance $G$, which leads to a suppression of $S \propto G^{-1}$ everywhere else. 
When $\Gamma_{\mathrm{CPS}} < \Gamma_S$, the $p$-wave case can be well distinguished from the $s$-wave case with the aid of the phase difference $\phi$.
Due to the small conductance, the Seebeck coefficient at $\phi = \pi$ in the $s$-wave case is one order of magnitude larger [orange and black solid curves in Fig.~\ref{fig:figure3}(a)] than the similar topological case [orange and black dashed curves in Fig.~\ref{fig:figure3}(b)].
At $\phi = 0$ the topological case shows a huge Seebeck response [orange and black dashed curves in Fig.~\ref{fig:figure3}(a)] for the same reason, while the corresponding conventional case is one order of magnitude smaller [orange and black solid curves in Fig.~\ref{fig:figure3}(b)].

Note that $S$ in Fig.~\ref{fig:figure3}(a) is solely determined by the EC transmission function $\mathcal{T}_{RL}^{ee}$, as LAR and CAR are absent for these phase differences at the symmetric choice of the direct couplings to the superconductors [cf.~Fig.~\ref{fig:figure1}] and remain negligible for small asymmetries.
However, LAR is not negligible in the conductance at other phases [i.e., Fig.~\ref{fig:figure3}(b)] and it is dominant in the $p$-wave case due to the MZM, as discussed in detail in Appendix~\ref{sec:appendixE}.

\subsection{Effect of the phase difference and CPS}
\begin{figure}
	\centering
	\includegraphics{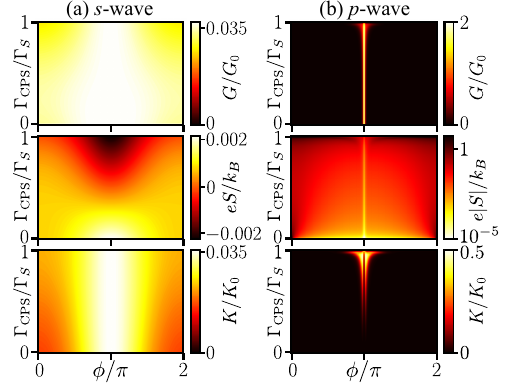}
	\caption{
		Electrical conductance $G$, Seebeck coefficient $S$, and thermal conductance $K$ for varying $\phi$ and $\Gamma_{\mathrm{CPS}}$.
		(a) $s$-wave case. (b) $p$-wave case. 
		$|S|$ in panel (b) is plotted on a logarithmic scale and is in the range $(eS/k_B) \in [-5.27,0.03]$.
		Parameters: $w = 3\Delta$,
		$\Gamma_N = \Delta / 5$, 
		$\Gamma_{S} = \Delta$,
		$\varepsilon_L = \varepsilon_R = -\Delta / 2$, 
		$k_B T = 10^{-3}\Delta$.
	}
	\label{fig:figure4}
\end{figure}
As we now show, measuring the thermoelectric coefficients in combination with tuning the applied phase difference across the JJ allows us to identify the presence of MZMs.
For this purpose, we show in Fig.~\ref{fig:figure4} the electrical conductance $G$, the Seebeck coefficient $S$, and the thermal conductance $K$ for different JJs.
As a first observation in the $s$-wave case [Fig.~\ref{fig:figure4}(a)], $G$, $S$, and $K$ show only an overall weak signal that is largest at $\phi = \pi$.
While both $G$ and $K$ do not significantly change on a qualitative level for different $\Gamma_{\mathrm{CPS}}$, $S$ shows a sign change around $\phi \approx \pi$ as $\Gamma_{\mathrm{CPS}}$ is increased. 
In contrast, the magnitude of the signals in the $p$-wave case [Fig.~\ref{fig:figure4}(b)] are much larger compared to the $s$-wave case.
In particular, $G$ shows the expected quantized conductance peak from LAR at the MZM at $\phi = \pi$ with $G = 2 G_0$ \cite{Yeyati_PRB_16} that is also robust against increasing $\Gamma_\mathrm{CPS}$, which is shown in Fig.~\ref{fig:figureS4}(a) in Appendix~\ref{sec:appendixE}.
At $\Gamma_\mathrm{CPS} = \Gamma_S$, $G$ is generally not quantized, but drops to $G = G_0$ in the case of symmetric couplings and equal dot levels $\varepsilon_L = \varepsilon_R$.
Furthermore, $K$ only shows a significant signal around phase differences close to $\phi \approx \pi$ for a large range of $\Gamma_{\mathrm{CPS}} \gtrsim 0.3 \, \Gamma_S $ that increases for increasing $\Gamma_{\mathrm{CPS}}$.
This behavior is a consequence of the violation of the Wiedemann-Franz law due to the presence of LAR; see Eq.~\eqref{eq:equation4} and Fig.~\ref{fig:figureS3}.
Although $K$ is zero at exactly $\phi = \pi$ for a large range of $\Gamma_{\mathrm{CPS}}$, it increases to the quantized value of $K = K_0/2$ in the case of symmetric coupling and equal dot energy, as shown in Fig.~\ref{fig:figureS4}(b) in Appendix~\ref{sec:appendixE}, which was also reported in Ref.~\cite{Bauer2021} for a continuous setup.
In addition, $S$ takes large values for small $\Gamma_{\mathrm{CPS}}$ at $\phi \approx 0$ due to the vanishingly small electrical conductance. 
This region [dark red area for $S$ in Fig.~\ref{fig:figure4}(b)] also extends to larger phase differences at large $\Gamma_{\mathrm{CPS}} \lesssim \Gamma_S$.

\section{Conclusion}
In the four-terminal setup shown in Fig.~\ref{fig:figure1}, we have investigated how ABS mediate the thermoelectric transport between two normal-metal contacts transverse to a Josephson junction.
In contrast to previous claims \cite{Lopez2014,Weymann2017,RamosAndrade2016,Ricco2018,Sun2021,Valentini2015}, we have shown that an inverted Seebeck coefficient is not a unique feature of MZMs and that ABS coexisting with resonant states [Fig.~\ref{fig:figure2}] or the strength of CPS in multi-dot systems also induce sign changes even in conventional JJs [Figs.~\ref{fig:figure3} and \ref{fig:figure4}(a)]. 

Moreover, we have compared the linear thermoelectric transport coefficients between conventional and topological JJs [Fig.~\ref{fig:figure4}]. 
In general, the signals in the conventional case are expected to be one to three orders of magnitude smaller than for the topological case. 
In particular, both the electrical and thermal conductance are strongly amplified at $\phi \approx \pi$ in the presence of MZMs, while the Seebeck coefficient is large around $\phi \approx 0$ due to the vanishingly small conductance. 
In addition, we have shown that the Wiedemann-Franz law is always strongly violated in the $p$-wave case around $\phi \approx \pi$, while it is always satisfied in the $s$-wave case.
This should represent a phase-sensitive and robust detection criterion for MZMs in such devices. 
As Cooper pair splitter experiments have been successfully performed and transverse geometries are in experimental reach in multiterminal setups, we believe that our theoretical proposal is ready to be implemented with current state of the art technology.

There are still numerous open questions in these types of systems, e.g., how these results connect to larger-scale continuous junctions or how the ABS in a mixed $s$- and $p$-wave situation influence the thermoelectric properties.
Furthermore, it will be interesting to investigate whether quasi-Majorana zero modes and the resulting Andreev bound states in nanowire-based Josephson junctions in the trivial regime can be distinguished from real MZMs in the topological regime with the help of transverse phase-dependent thermoelectric transport \cite{Kells2012,Moore2018,Zeng2022}.

\begin{acknowledgments}
We acknowledge Martin Stehno and Yi-Ju Ho for fruitful discussions. 
We acknowledge funding by the Deutsche Forschungsgemeinschaft (DFG, German Research Foundation) through SFB1170 ToCoTronics, Project-ID 258499086, through the Würzburg-Dresden Cluster of Excellence on Complexity and Topology in Quantum
Matter – ct.qmat (EXC2147, Project-ID 390858490).
We also gratefully acknowledge the Gauss Centre for Supercomputing e.V. (\url{www.gauss-centre.eu}) for funding this project by providing computing time on the GCS Supercomputer SuperMUC-NG at Leibniz Supercomputing Centre (\url{www.lrz.de}).
L.W.M. acknowledges support from the Free State of Bavaria for the Institute for Topological Insulators.
\end{acknowledgments}

\appendix

\section{Boundary Green's functions of the four terminals}\label{sec:appendixA}
The left $(L)$, right $(R)$, top $(T)$, and bottom $(B)$ leads in the four-terminal junction that is sketched in Fig.~1(a) of the main text are all modelled by semi-infinite tight-binding chains, generally described by the set of Hamiltonians
\begin{align}
	H_\mathrm{leads} = H_{L}^{(n)} + H_{R}^{(n)} + H_{T}^{(s/p)} + H_{B}^{(s/p)}.
\end{align}
We assume that the left and right terminals are normal metallic electrodes (superscript $n$), while the top and bottom leads are superconductors of either conventional $s$-wave or topological $p$-wave type (superscript $s$ or $p$). 
The Hamiltonian $H_{\ell}^{(s)}$ for the conventional $s$-wave superconductors is given by the tight-binding chain
\begin{widetext}
\begin{align}
	\label{eq:swaveHamiltonian}
	H_{\ell}^{(s)} &= \sum_{j} 
	\Bigl[ 
	-t \sum_\sigma ( c_{\ell,j,\sigma}^\dagger c_{\ell,j+1,\sigma} 
	+ c_{\ell,j+1,\sigma}^\dagger c_{\ell,j,\sigma} )
	+ \Delta e^{i\varphi_\ell} c_{\ell,j,\uparrow}^\dag c_{\ell,j,\downarrow}^\dag 
	+ \Delta e^{-i\varphi_\ell} c_{\ell,j,\downarrow}  c_{\ell,j,\uparrow}
	\Bigr]
	\nonumber \\
	&= \sum_{k} (c_{\ell,k,\uparrow}^\dag, c_{\ell,-k,\downarrow}) \Bigl( -2 t \cos(k) \tau_3 + \Delta e^{i\varphi_\ell \tau_3} \tau_1 \Bigr)  
	\begin{pmatrix}
		c_{\ell,k,\uparrow}\\ c_{\ell,-k,\downarrow}^\dag
	\end{pmatrix},
\end{align}
where $c_{\ell,j,\sigma} = \sum_k c_{\ell, k,\sigma} e^{- i j k} / \sqrt{N}$ annihilates and $c_{\ell,j,\sigma}^{\dagger} = \sum_k c^{\dag}_{\ell, k,\sigma} e^{i j k} / \sqrt{N}$ creates an electron of spin $\sigma \in \{ \uparrow, \downarrow \}$ on site $j$ in terminal $\ell = T,B$.
Furthermore, $t > 0$ is the hopping parameter that is related to the lattice spacing defining the effective bandwidth in the model, the number of sites is $N \to \infty$, $\Delta \geq 0$ is the superconducting order parameter, and $\varphi_\ell$ is its phase.
We also introduced a set of Pauli matrices $\tau_1$, $\tau_2$, and $\tau_3$ in Nambu space, with $\tau_0$ being the identity.

If the top and bottom superconducting terminals are of topological $p$-wave type, we describe them by spinless Kitaev chains (at zero onsite potential) with the Hamiltonian \cite{kitaevchain,Yeyati_PRB_16}
\begin{align}
	\label{eq:pwaveHamiltonian}
	H_\ell^{(p)} &=
	\frac{1}{2} \sum_j \left[ 
	-t (c^\dag_{\ell, j} c^{\phantom \dag}_{\ell, j+1} +  c^\dag_{\ell, j+1} c_{\ell, j} )
	+ \Delta e^{i\varphi_\ell} c^\dag_{\ell, j} c^\dag_{\ell, j+1} + \Delta e^{- i\varphi_\ell} c_{\ell, j+1}^{\phantom \dag} c_{\ell, j}^{\phantom \dag}   \right] 
	\nonumber \\
	&= 
	\frac{1}{2} \sum_k ( c^\dag_{\ell,k} , c_{\ell,-k} )
	\Bigl( - t \cos(k) \tau_3 + \Delta  \sin(k) e^{i\varphi_\ell \tau_3} \tau_2 \Bigr)
	\begin{pmatrix}
		c_{\ell,k} \\ c^\dag_{\ell,-k}
	\end{pmatrix}
	.
\end{align}
\end{widetext}
Note that the topological $p$-wave terminal is considered to be spinless and, hence, $c_{\ell,j}^{(\dagger)}$ does not depend on spin. 
In both the $s$-wave and $p$-wave case, the normal-metal Hamiltonian $H_{\ell}^{(n)}$ ($\ell = L,R$) follows from $H_{\ell}^{(s)}$ for $\Delta = 0$.

Following standard Green's function techniques 
\cite{Cuevas1996,cuevas2nd,Yeyati_PRB_16,Yeyati_PRB_20}, we obtain the bare matrix boundary Green's functions [in their respective basis defined in Eqs.~\eqref{eq:swaveHamiltonian} and \eqref{eq:pwaveHamiltonian}] for the semi-infinite terminals as 
\begin{subequations}
	\label{eq:BGFs}
	\begin{align}
		g^{(n)}_\ell(z) &= - \frac{i}{t} \mathrm{sgn}(\mathrm{Im}(z))\tau_0,
		\qquad 
		\ell = L,R ,
		\label{eq:normalBGF}
		\\
		g_\ell^{(s)}(z) &= \frac{- 
			(z \tau_0 + \Delta e^{i \varphi_\ell \tau_3} \tau_1) 
		}{t \sqrt{ \Delta^2 - z^2 } } ,
		\qquad 
		\ell = T,B ,
		\label{eq:conventionalBGF}
		\\
		g_\ell^{(p)}(z) &= \frac{ \sqrt{ \Delta^2 - z^2 } \tau_0
			+ 
			\Delta_\ell e^{i \varphi_\ell \tau_3} \tau_1
		}{t z } ,
		\quad 
		\ell = T,B ,
		\label{eq:topologicalBGF}
	\end{align}
\end{subequations}
where $z = \varepsilon + i \eta$, with energy $\varepsilon \in \mathbb{R}$ and a small Dynes parameter $\eta \to 0^{\pm}$ \cite{Dynes1978}. 
In our numerical calculations we used $|\eta| = 10^{-3}\Delta$ for Figs.~\ref{fig:figure2}\textendash\ref{fig:figure3} and $|\eta| = 10^{-5}\Delta$ for Fig.~\ref{fig:figure4} in the main text.
The retarded and advanced Green's functions are defined for $\eta > 0$ and $\eta < 0$, respectively.
We further assumed that the band parameter $t$, which also plays the role of the inverse normal-metal density of states at the Fermi energy, is the largest energy scale in the total system (wide-band approximation) \cite{cuevas2nd}.
Finally, note that the order parameter in the topological case reads $\Delta_T = \Delta$ and $\Delta_B = - \Delta$ due to the $p$-wave nature of the superconducting pairing \cite{Yeyati_PRB_16}.

In the wide-band limit, the normal-metal Green's function $g^{(n)}_\ell(z)$ in Eq.~\eqref{eq:normalBGF} is constant in energy, while the conventional superconductor described by $g_\ell^{(s)}(z)$ in Eq.~\eqref{eq:conventionalBGF} shows the characteristic BCS singularities at the energies $\varepsilon = \pm \Delta$.
In contrast, the topological superconductor described by $g_\ell^{(p)}(z)$ in Eq.~\eqref{eq:topologicalBGF} hosts a single Majorana state at $\varepsilon = 0$.

\section{Dressed central double-quantum dot and couplings}\label{sec:appendixB}
As explained in the main text, the central quantum-dot system is described by an interconnected double-quantum dot.
In the spinful (i.e., $s$-wave) case, the two quantum dots are considered to be spin-degenerate and noninteracting, as described by the Hamiltonian 
\begin{widetext}
\begin{align}
	H_{\mathrm{QD}}^{(s)}
	&= \sum_{\alpha} \sum_\sigma \varepsilon_\alpha d_{\alpha\sigma}^\dagger d_{\alpha\sigma}
	-
	w \sum_\sigma (d_{L\sigma}^\dagger d_{R\sigma} + d_{R\sigma}^\dagger d_{L\sigma} )
	= 
	(d_{L\uparrow}^\dag , d_{L\downarrow} , d_{R\uparrow}^\dag , d_{R\downarrow} )
	\underbrace{ \left[  \begin{pmatrix}
			\varepsilon_L  & -w  \\
			-w  &\varepsilon_R  
		\end{pmatrix} \otimes \tau_3 \right] }_{ \hat{H}_{\mathrm{QD}} }
	\begin{pmatrix}
		d_{L\uparrow} \\
		d_{L\downarrow}^\dag \\
		d_{R\uparrow} \\
		d_{R\downarrow}^\dag 
	\end{pmatrix} ,
\end{align}
where $d_{\alpha\sigma}^{(\dagger)}$ annihilates (creates) an electron of spin $\sigma \in \{ \uparrow, \downarrow \}$ at the onsite energy $\varepsilon_\alpha \in \mathbb{R}$ of QD $\alpha \in \{L,R\}$ and $w \geq 0$ is the coupling strength between the two quantum dots.
For the spinless $p$-wave case, we get the similar result
\begin{align}
	H_{\mathrm{QD}}^{(p)}
	&= \sum_{\alpha} \varepsilon_\alpha d_{\alpha}^\dagger d_{\alpha}
	-
	w (d_{L}^\dagger d_{R} + d_{R}^\dagger d_{L} )
	= 
	\frac{1}{2} (d_{L}^\dag , d_{L} , d_{R}^\dag , d_{R} )
	\hat{H}_{\mathrm{QD}}
	\begin{pmatrix}
		d_{L} \\
		d_{L}^\dag \\
		d_{R} \\
		d_{R}^\dag 
	\end{pmatrix} .
\end{align}
As sketched in Fig.~\ref{fig:figure1}(a) in the main text, the left (right) normal-metal terminal is only coupled to the left (right) quantum dot, while the top and bottom superconducting terminals are each coupled to both quantum dots simultaneously.
The full coupling Hamiltonian between the terminals and the central quantum dot system reads
\begin{align}
	H_\mathrm{tunnel}^{(s)}
	=
	\sum_{\ell = L,R,T,B}
	H_{\ell D}^{(s)},
\end{align}
where in the $s$-wave case 
\begin{subequations}
	\begin{align}
		H_{\ell D}^{(s)} &= \sum_{ k \sigma} (-t_\ell d_{\ell\sigma}^\dagger c_{\ell k \sigma} + \mathrm{H.c.}) 
		=
		\sum_{k} 
		\left( 
		(d_{\ell\uparrow}^\dagger , d_{\ell\downarrow})
		V_\ell 
		\begin{pmatrix}
			c_{\ell k \uparrow} \\ c_{\ell,-k,\downarrow}^\dagger
		\end{pmatrix}
		+
		\mathrm{H.c.} \right),
		\qquad \ell = L,R ,
		\\
		H_{\ell D}^{(s)} &= \sum_{\alpha = L,R} \sum_{ k \sigma} (-t_{\ell\alpha} d_{\alpha \sigma}^\dagger c_{\ell k \sigma} + \mathrm{H.c.}) 
		=
		\sum_{k} 
		\left( 
		(d_{\alpha\uparrow}^\dagger , d_{\alpha\downarrow})
		V_{\ell\alpha} 
		\begin{pmatrix}
			c_{\ell k \uparrow} \\ c_{\ell,-k,\downarrow}^\dagger
		\end{pmatrix}
		+
		\mathrm{H.c.} \right), \qquad \ell = T,B ,
	\end{align}
\end{subequations}
and in the $p$-wave case
\begin{subequations}
	\begin{align}
		H_{\ell D}^{(p)} &= \sum_{ k} (-t_\ell d_{\ell}^\dagger c_{\ell k } + \mathrm{H.c.}) 
		=
		\frac{1}{2}
		\sum_{k} 
		\left( 
		(d_{\ell}^\dagger , d_{\ell})
		V_\ell 
		\begin{pmatrix}
			c_{\ell k } \\ c_{\ell,-k}^\dagger
		\end{pmatrix}
		+
		\mathrm{H.c.} \right),
		\qquad \ell = L,R ,
		\\
		H_{\ell D}^{(p)} &= \sum_{\alpha = L,R} \sum_{ k } (-t_{\ell\alpha} d_{\alpha }^\dagger c_{\ell k } + \mathrm{H.c.}) 
		=
		\frac{1}{2} \sum_{\alpha = L,R} \sum_{ k }
		\left( 
		(d_{\alpha}^\dagger , d_{\alpha})
		V_{\ell\alpha} 
		\begin{pmatrix}
			c_{\ell k } \\ c_{\ell,-k}^\dagger
		\end{pmatrix}
		+
		\mathrm{H.c.} \right), \qquad \ell = T,B .
	\end{align}
\end{subequations}
We also defined the coupling matrices $V_\ell = -t_\ell \tau_3$ and $V_{\ell\alpha} = -t_{\ell\alpha} \tau_3$, where $t_\ell, t_{\ell\alpha} \geq 0$.
We use the Dyson equation $G = g + g\Sigma G$ to calculate the dressed retarded and advanced Green's function $\hat{G}_{CC}$ of the central system as
\begin{align}
	\hat{G}_{CC}(z) &= \begin{pmatrix}
		G_{LL}(z) & G_{LR}(z) \\ G_{RL}(z) & G_{RR}(z)
	\end{pmatrix}
	= (z \hat{\mathbbm{1}}_{4} - \hat{H}_{\mathrm{QD}} - \hat{\Sigma}(z) )^{-1}, 
	\label{eq:centralGF}
\end{align}
where $\hat{\mathbbm{1}}_{4}$ is a $4\times 4$ unit matrix, $z = \varepsilon + i \eta$ [cf.~Eq.~\eqref{eq:BGFs}], and the self-energy matrix is given by 
\begin{align}
	\hat{\Sigma}(z) = 
	\begin{pmatrix}
		\Sigma_{LL}(z) & \Sigma_{LR}(z)
		\\
		\Sigma_{RL}(z) & \Sigma_{RR}(z)
	\end{pmatrix} ,
\end{align}
where 
\begin{align}
	\Sigma_{\ell\ell'}(z) &= \delta_{\ell \ell'} V_\ell g_\ell^{(n)}(z) V_\ell + \sum_{\alpha = T,B} V_{\alpha,\ell} g_\alpha^{(s/p)}(z)  V_{\alpha,\ell'} 
\end{align}
and $\delta_{\ell \ell'}$ is the Kronecker-$\delta$,
which simplifies to
\begin{subequations}
	\label{eq:selfenergyelements}
	\begin{align}
		\Sigma_{\ell\ell'}(z) &= 
		- i \delta_{\ell \ell'} \frac{t_{\ell}^2 }{t} \mathrm{sgn}(\mathrm{Im}(z)) \tau_0
		+ \sum_{\beta = T,B} \frac{- t_{\beta\ell} t_{\beta\ell'} (z \tau_0 - \Delta e^{i \varphi_\beta \tau_3} \tau_1)}{ t \sqrt{\Delta^2 - z^2} }
		\qquad\qquad \text{($s$-wave)} ,
		\\
		\Sigma_{\ell\ell'}(z) &= 
		- i \delta_{\ell \ell'} \frac{t_{\ell}^2 }{t} \mathrm{sgn}(\mathrm{Im}(z)) \tau_0
		+ \sum_{\beta = T,B} \frac{t_{\beta\ell} t_{\beta\ell'} (\sqrt{\Delta^2 - z^2} \tau_0 - \Delta_\beta e^{i \varphi_\beta \tau_3} \tau_1)}{ t z }
		\qquad \text{($p$-wave)} .
	\end{align}
\end{subequations}
In Eq.~\eqref{eq:selfenergyelements}, the origin of the first term is the coupling of the left (right) quantum dot to the left (right) normal-metal terminal, while the second term arises due to the coupling of both dots to both top and bottom superconducting leads.
In addition, the coupling of both dots to a single superconductor adds nonlocal terms $\Sigma_{LR}(z)$ and $\Sigma_{RL}(z)$ that lead Cooper pair splitting (CPS) into two electrons.

As introduced in the main text, the normal-metal terminals are coupled to the double quantum dot with the effective couplings $\Gamma_\ell = t_\ell^2 / t \geq 0$ ($\ell = L,R$), while the superconducting terminals are locally coupled to the double quantum dot with the effective couplings $\Gamma_{\beta,\ell} = t_{\beta,\ell}^2 / t \geq 0$ ($\beta = T,B$).
In addition, the nonlocal couplings $\Gamma_{\beta,\mathrm{NL}} = t_{\beta,L} t_{\beta,R}/t \geq 0$ control the CPS.
Since in general experimental situations we have $\Gamma_{\beta,\mathrm{NL}} < \sqrt{\Gamma_{\beta,L}\Gamma_{\beta,R}}$, we consider $\Gamma_{\beta,\mathrm{NL}}$ as an independent parameter in our theory. 
In the main text, we choose symmetric couplings for simplicity: $\Gamma_N := \Gamma_L = \Gamma_R$, $\Gamma_S := \Gamma_{T,L} = \Gamma_{T,R} = \Gamma_{B,L}= \Gamma_{B,R}$, and $\Gamma_{\mathrm{CPS}} := \Gamma_{T,\mathrm{NL}} = \Gamma_{B,\mathrm{NL}}$.

Note that for the symmetric choice of couplings and the assumption of equal magnitude of the pairings, the self-energies simplify to 
\begin{subequations}
	\label{eq:absenceOfSC}
	\begin{align}
		\Sigma_{\ell\ell'}(z) &= 
		- i \delta_{\ell \ell'} \frac{t_{\ell}^2 }{t} \mathrm{sgn}(\mathrm{Im}(z)) \tau_0
		-  \frac{ t_{S,\ell} t_{S,\ell'} }{ t \sqrt{\Delta^2 - z^2} } \left(  2 z \tau_0 
		- \Delta [ e^{i \varphi_T \tau_3}  
		+  e^{i \varphi_B \tau_3}  ] \tau_1 \right) 
		\qquad\qquad \text{($s$-wave)} .
		\\
		\Sigma_{\ell\ell'}(z) &= 
		- i \delta_{\ell \ell'} \frac{t_{\ell}^2 }{t} \mathrm{sgn}(\mathrm{Im}(z)) \tau_0
		+ \frac{t_{S,\ell} t_{S,\ell'} }{ t z } 
		\left(2 \sqrt{\Delta^2 - z^2} \tau_0 - \Delta [e^{i \varphi_T \tau_3}  - e^{i \varphi_B \tau_3}  ] \tau_1\right)
		\qquad \text{($p$-wave)} .
	\end{align}
\end{subequations}
\end{widetext}
Hence, there is no effect of superconductivity at a phase difference $\phi = \varphi_B - \varphi_T = \pi$ in the $s$-wave case, while the same effect happens at a phase difference $\phi = \varphi_B - \varphi_T = 0$ in the $p$-wave case, which leads to the absence of both local and crossed Andreev reflection at these phase differences.

\section{Transmission functions for charge and heat currents}\label{sec:appendixC}
In the following, we focus on the electric current $I_R$ and the heat current $J_R$ in the right normal-metal contact that are generated by applying a voltage or thermal bias to the right contact.
Since we do not consider a voltage bias between the superconducting terminals, these currents will be stationary (i.e., time-independent).
Hence, by starting from the definition of the currents $I_R = - e \langle d N_R / dt \rangle$ and $J_R = \langle d (H_R-\mu_R N_R) / dt \rangle$, with the elementary charge $e>0$ and $N_R$, $H_R$, and $\mu_R$ being the particle number operator, the Hamiltonian, and the chemical potential in the right contact, respectively, and by using a Fourier transformation to energy space, we get
\begin{multline}
	\begin{pmatrix}
		I_R \\ J_R
	\end{pmatrix}
	= \frac{1}{2h} \int_{-\infty}^\infty \mathrm{Tr}\Biggl\{ 
	\begin{pmatrix}
		e \tau_3 \\ \varepsilon \tau_0 - \mu_R \tau_3
	\end{pmatrix}
	\\
	\label{eq:thermoelectricCurrent}
	\times \Bigl[ V_{RC} \, G_{CR}^<(\varepsilon) -  G_{RC}^<(\varepsilon) \, V_{CR} \Bigr] \Biggr\} \, d\varepsilon .
\end{multline}
Note that the trace is taken over the particle-hole (Nambu) degrees of freedom.
Furthermore, $h$ is Planck's constant, $G_{RC}^<(\varepsilon)$ and $G_{CR}^<(\varepsilon)$ are lesser dressed Green's functions, and $V_{\ell C} = V_{C \ell} = -t_\ell \tau_3$ are hoppings between the central quantum dot and the terminals. 
We use the Dyson equations for the lesser Green's function \cite{cuevas2nd}, 
\begin{subequations}
	\label{eq:dysonLesser1}
	\begin{align}
		G_{CR}^< &= G_{CC}^< V_{CR} g_{R}^a + G^r_{CC} V_{CR} g_R^< ,
		\\
		G_{RC}^< &= g_{R}^< V_{RC} G_{CC}^a + g^r_{R} V_{RC} G_{CC}^< ,
	\end{align}
\end{subequations}
where the superscripts $r$ and $a$ are for the retarded and advanced Green's functions, respectively.
Using the general relation $G^< - G^> = G^a - G^r$ that holds for both dressed and bare Green's functions \cite{cuevas2nd}, Eq.~\eqref{eq:thermoelectricCurrent} becomes
\begin{multline}
	\begin{pmatrix}
		I_R \\ J_R
	\end{pmatrix}
	= \frac{1}{2h} \int_{-\infty}^\infty \mathrm{Tr}\Biggl\{ 
	\begin{pmatrix}
		e \tau_3 \\ \varepsilon \tau_0 - \mu_R \tau_3
	\end{pmatrix}
	\\
	\times
	\Bigl[ 
	V_{RC} G^>_{CC} V_{CR} g_R^<
	-
	V_{RC} G_{CC}^< V_{CR} g_R^>
	\Bigr] \Biggr\} \, d\varepsilon .
\end{multline}
Finally, we use the symmetric version of the Dyson equation for the dressed lesser and greater Green's function \cite{cuevas2nd}
\begin{align}
	\label{eq:dysonLesser2}
	G_{CC}^{<,>} = \sum_{\ell = L,R} G^r_{CC} V_{C\ell} g^{<,>}_\ell V_{\ell C} G^a_{CC} 
\end{align}
to arrive at
\begin{multline}
	\begin{pmatrix}
		I_R \\ J_R
	\end{pmatrix}
	= \frac{1}{2h} \sum_{\ell = L,R}  \int_{-\infty}^\infty \mathrm{Tr}\Biggl\{ 
	\begin{pmatrix}
		e \tau_3 \\ \varepsilon \tau_0 - \mu_R \tau_3
	\end{pmatrix}
	 \\
	\times V_{RC} G^r_{CC} V_{C\ell} 
	\biggl[ 
	g^{>}_\ell V_{\ell C} G^a_{CC}  V_{CR} g_R^<
	 \\
	 -
	g^{<}_\ell V_{\ell C} G^a_{CC}  V_{CR} g_R^>
	\biggr] \Biggr\} \, d\varepsilon .
	\label{eq:thermoelectricCurrent2}
\end{multline}
Note that we neglect the current contributions from and into the superconductors in Eq.~\eqref{eq:dysonLesser2} since we are only interested in linear response around zero energy. 
The uncoupled normal terminals are in equilibrium, in which the bare lesser and greater Green's function is given by 
\begin{subequations}
	\label{eq:bareLesserGreater}
	\begin{align}
		\label{eq:bareLesser}
		g_\ell^< &= (g_\ell^a - g_\ell^r) \begin{pmatrix}
			f_\ell^e & 0 \\
			0 & f_\ell^h
		\end{pmatrix} 
		\stackrel{\eqref{eq:normalBGF}}{=}
		\frac{2i}{t}\begin{pmatrix}
			f_\ell^e & 0 \\
			0 & f_\ell^h
		\end{pmatrix} ,
		\\
		g_\ell^> &= -(g_\ell^a - g_\ell^r) \begin{pmatrix}
			1-f_\ell^e & 0 \\
			0 & 1-f_\ell^h
		\end{pmatrix} 
		\nonumber \\
		&\stackrel{\eqref{eq:normalBGF}}{=}
		- \frac{2i}{t}\begin{pmatrix}
			1- f_\ell^e & 0 \\
			0 & 1- f_\ell^h
		\end{pmatrix} 
			\label{eq:bareGreater}.
	\end{align}
\end{subequations}
Here, $f^{e,h}_\ell(\varepsilon) = 1/[1+e^{(\varepsilon\mp\mu_\ell)/(k_B T_\ell)}]$ is the Fermi function for electrons and holes, respectively, of the left and right terminals, which are at chemical potential $\mu_\ell$ and temperature $T_\ell$, and $k_B$ is Boltzmann's constant.
Substituting Eq.~\eqref{eq:bareLesserGreater} into  Eq.~\eqref{eq:thermoelectricCurrent2} and taking the trace over particle-hole space, we finally get
\begin{widetext}
\begin{align}
	\label{eq:thermoelectricCurrent3}
	\begin{pmatrix}
		I_R \\ J_R
	\end{pmatrix}
	&= \frac{1}{2h} \int_{-\infty}^\infty  
	\begin{pmatrix}
		e  
		(
		\mathcal{T}_{RL}^{ee} (f_R^e - f_L^e) 
		+  \mathcal{T}_{RL}^{eh}  (f_R^e - f_L^h) 
		- \mathcal{T}_{RL}^{he} (f_R^h - f_L^e) 
		- \mathcal{T}_{RL}^{hh} (f_R^h - f_L^h) 
		) 
		\\ 
		(\varepsilon-\mu_R) [\mathcal{T}_{RL}^{ee} (f_R^e-f_L^e)
		+ \mathcal{T}_{RL}^{eh} (f_R^e-f_L^h) ]
		+ (\varepsilon+\mu_R) [\mathcal{T}_{RL}^{he} (f_R^h-f_L^e)
		+ \mathcal{T}_{RL}^{hh} (f_R^h-f_L^h) ]
	\end{pmatrix} d\varepsilon 
	\nonumber \\
	&\quad + \frac{1}{2h} \int_{-\infty}^\infty  
	\begin{pmatrix}
		e 
		(  \mathcal{T}_{RR}^{eh} + \mathcal{T}_{RR}^{he}  ) 
		\\ 
		(\varepsilon - \mu_R ) \mathcal{T}_{RR}^{eh} 
		- (\varepsilon + \mu_R ) \mathcal{T}_{RR}^{he} 
	\end{pmatrix} ( f_R^e - f_R^h ) \, d\varepsilon ,
\end{align}
\end{widetext}
where we defined the transmission functions
\begin{align}
	\label{eq:transmissionFunctions}
	\mathcal{T}_{\ell_1 \ell_2}^{\tau_1 \tau_2}(\varepsilon) = 4 \Gamma_{\ell_1} \Gamma_{\ell_2} \, G^r_{\ell_1\tau_1, \ell_2  \tau_2} (\varepsilon) \, G^a_{\ell_2  \tau_2 , \ell_1 \tau_1 }(\varepsilon) .
\end{align}
The special case for $\Gamma_L = \Gamma_R = \Gamma_N$ is presented in Eq.~\eqref{eq:equation1} of the main text.
\begin{figure*}
	\centering
	\includegraphics{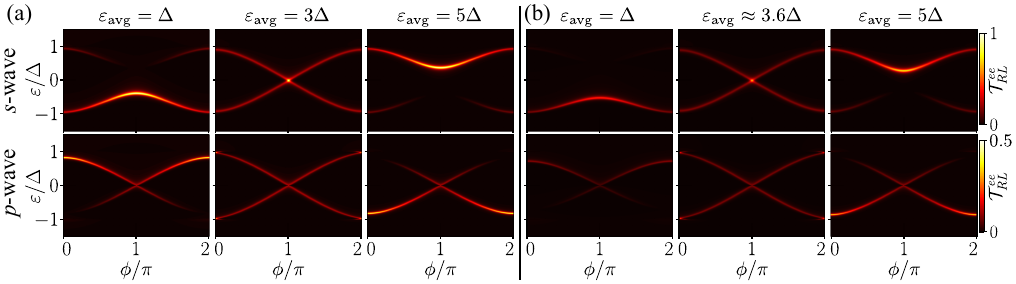}
	\caption{
		EC transmission function to show the "inversion" between electron or hole-dominated transport. 
		The inversion is always happening in both systems when the gap is closed in the $s$-wave case. 
		Definition: $\varepsilon_\mathrm{avg/diff} = (\varepsilon_L \pm \varepsilon_R)/2$.
		(a) For $\varepsilon_\mathrm{diff} = 0$, the crossover takes place at $\varepsilon_\mathrm{avg} = w$.
		(b) For $\varepsilon_\mathrm{diff} = 2\Delta$, the crossover takes place at $\varepsilon_\mathrm{avg} \approx 3.6 \Delta$.
		Common parameters: 
		$w = 3\Delta$,
		$\eta = 10^{-3}\Delta$,
		$k_B T = 10^{-2}\Delta$,
		$\Gamma_N = \Delta/5$,
		$\Gamma_S = \Delta$,
		$\Gamma_{\rm CPS} = \Gamma_S$.
	}
	\label{fig:figureS1}
\end{figure*}
Using particle-hole symmetry, which states $f^{\tau}_\ell(\varepsilon) = 1-f^{\bar{\tau}}_\ell(-\varepsilon)$ and $\mathcal{T}_{\ell_1 \ell_2}^{\tau_1 \tau_2}(\varepsilon) = \mathcal{T}_{\ell_1 \ell_2}^{\bar{\tau}_1 \bar{\tau}_2}(-\varepsilon)$ with $\bar{\tau} = e,h$ if $\tau = h,e$, we can further simplify Eq.~\eqref{eq:thermoelectricCurrent3} to obtain 
\begin{multline}
	\label{eq:thermoelectricCurrent4}
	\begin{pmatrix}
		I_R \\ J_R
	\end{pmatrix}
	= \frac{1}{h} \int_{-\infty}^\infty  
	\begin{pmatrix}
		e  \\  \varepsilon-\mu_R
	\end{pmatrix} 
	\Bigl[
	\mathcal{T}_{RL}^{ee} (f_R^e-f_L^e) 
	\\
	+ \mathcal{T}_{RL}^{eh} (f_R^e-f_L^h) 
	+ \mathcal{T}_{RR}^{eh} (f_R^e-f_R^h)
	\Bigr] 
	\, d\varepsilon .
\end{multline}
As explained in the main text, we consider a small voltage $\delta V$ and thermal bias $\delta T$ applied at the right normal-metal terminal, i.e.: $e \, \delta V \ll k_B T$ and $\delta T \ll  T$, respectively.
Hence, we choose $T_L = T$, $\mu_L = 0$, $T_R = T + \delta T$, and $\mu_R = e \, \delta V$.
By expanding the integrand in Eq.~\eqref{eq:thermoelectricCurrent4} to linear order in $\delta T$ and $\delta V$, we arrive at the linear Onsager relations presented in Eqs.~\eqref{eq:equation2}\textendash\eqref{eq:equation4} in the main text.

\section{Complementary figures}\label{sec:appendixD}
\begin{figure*}
	\centering
	\includegraphics{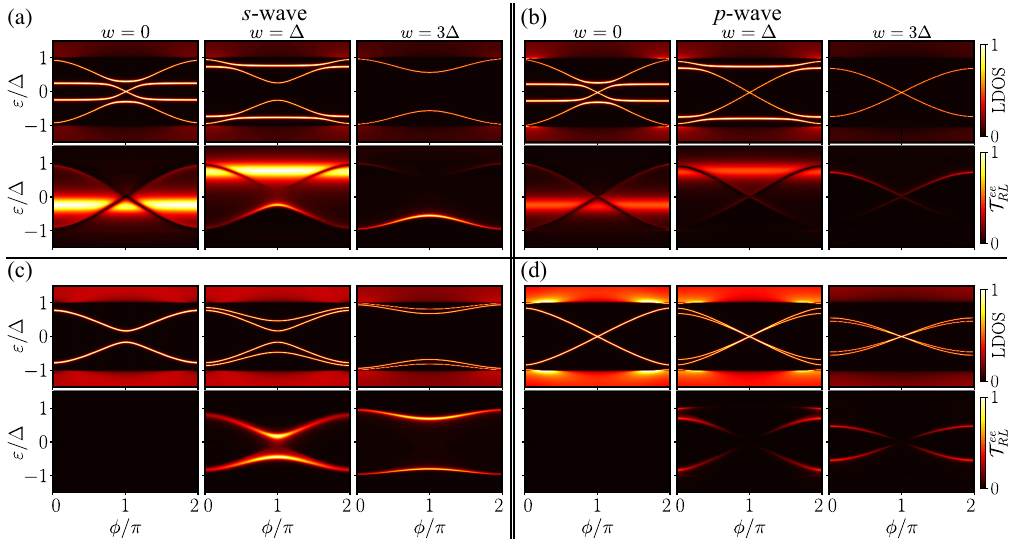}
	\caption{
		Complementary figures to Fig.~\ref{fig:figure2} of the main text of the LDOS and the EC transmission function.
		(a)-(b) Asymmetric quantum dot energies $\varepsilon_L = -\Delta/2$, $\varepsilon_R = 0$ and maximal CPS $\Gamma_{\mathrm{CPS}} = \Gamma_S$.
		(c)-(d) Symmetric quantum dot energies $\varepsilon_L = \varepsilon_R = -\Delta/2$ and no CPS $\Gamma_{\mathrm{CPS}} = 0$.
		Common parameters: 
		$\eta = 10^{-3}\Delta$,
		$k_B T = 10^{-2}\Delta$,
		$\Gamma_N = \Delta/5$,
		$\Gamma_S = \Delta$.
	}
	\label{fig:figureS2}
\end{figure*}
Figure \ref{fig:figureS1} shows the transition between the hole- and electron-dominated transport regimes by means of the EC transmission function $\mathcal{T}^{ee}_{RL}$. 
In general, the crossover between these two regimes takes place whenever the gap in the $s$-wave case is closed at $\phi = \pi$ (top row in Figure \ref{fig:figureS1}).
This also happens at the same parameter set in the $p$-wave case, although the gap is always closed due to its topological protection and the presence of a Majorana zero mode (MZM) at $\phi = \pi$.
Note that the EC transmission function is symmetric in energy $\varepsilon$ at the crossover point and does not contribute to the off-diagonal Onsager coefficients $\mathcal{L}_{12}$ and $\mathcal{L}_{21}$.
For a symmetric choice of QD energies $\varepsilon_L = \varepsilon_R$ [Fig.~\ref{fig:figureS1}(a)], the crossover takes place at $\varepsilon_\mathrm{avg} = w$.
The same closing of the gap happens also in the asymmetric case $\varepsilon_L \neq \varepsilon_R$ shown in Fig.~\ref{fig:figureS1}(b), which, however, shifts the crossover point to a different value of $\varepsilon_\mathrm{avg}$.

Figure \ref{fig:figureS2} shows the complementary cases to Fig.~\ref{fig:figure2} that were briefly mentioned in the main text.
In particular, in Figs.~\ref{fig:figureS2}(a) and \ref{fig:figureS2}(b) we show the LDOS and EC transmission function for small asymmetric values of the quantum dot energies, $\varepsilon_L \neq \varepsilon_R$.
Again, for $\Gamma_\mathrm{CPS} = \Gamma_S$, we observe the constant resonant line in the EC transmission function which we also see in the symmetric case, although now the states are hybridized. 
Furthermore, as shown in Figs.~\ref{fig:figureS2}(c) and \ref{fig:figureS2}(d), the constant resonant line disappears for weak CPS and a second pair of Andreev states emerges.

\section{Linear response electrical and thermal conductance at low temperature}\label{sec:appendixE}
The general elements of the Onsager matrix $\mathcal{L}$ are given in Eq.~\eqref{eq:equation3} in the main text.
Their low-temperature behavior is obtained by means of a Sommerfeld expansion of $\mathcal{L}_{mn}$, resulting in \cite{benenti2017fundamental}
\begin{subequations}
\begin{align}
	\mathcal{L}_{11} &\approx \mathcal{T}_{11}(0) ,
	\\  
	\mathcal{L}_{12}  &\approx \frac{\pi^2}{3} k_B T \left. \frac{\partial \mathcal{T}_{12}}{\partial \varepsilon}\right|_{\varepsilon=0}, 
	\\ 
	\mathcal{L}_{21}  &\approx \frac{\pi^2}{3} k_B T \left. \frac{\partial \mathcal{T}_{21}}{\partial \varepsilon}\right|_{\varepsilon=0} , 
	\\ 
	\mathcal{L}_{22}  &\approx \frac{\pi^2}{3}  \mathcal{T}_{22}(0) ,
\end{align}
\end{subequations}
where the general transmission functions $\mathcal{T}_{mn}$ are defined in Eq.~\eqref{eq:equation4} in the main text. 
Note that, although $\mathcal{T}_{21} \neq \mathcal{T}_{12}$ due to local Andreev reflection, we still have $\mathcal{L}_{12} = \mathcal{L}_{21}$ since the corresponding transmission function $\mathcal{T}_{RR}^{eh}$ is symmetric in energy and its derivative at zero energy is zero. 
From these equations, we get the low-temperature behavior of the electrical and thermal conductance as
\begin{subequations}
\label{supl_eq: conductances}
\begin{align}
	\frac{G}{G_0} &= \mathcal{L}_{11} \approx \mathcal{T}_{11}(0),  
	\\
	\frac{K}{K_0} &= \frac{3}{\pi^2} \, \frac{ \mathrm{det}(\mathcal{L}) }{ \mathcal{L}_{11} }
	\nonumber \\
	&\approx  
	\mathcal{T}_{22}(0)
	-
	\frac{\pi^2}{3} 
	\frac{  (k_B T)^2 }{ \mathcal{T}_{11}(0) } 
	\left. \frac{\partial \mathcal{T}_{12}}{\partial \varepsilon}\right|_{\varepsilon=0}
	\left. \frac{\partial \mathcal{T}_{21}}{\partial \varepsilon}\right|_{\varepsilon=0} 
	,
\end{align} 
\end{subequations}
resulting in $K \approx K_0 \, \mathcal{T}_{22}(0)$ at low temperature as long as the second term remains negligible. 
Since in general $\mathcal{T}_{11} \neq \mathcal{T}_{22}$ due to LAR, Eq.~\eqref{supl_eq: conductances} shows the violation of the WF law that is illustrated in Fig.~\ref{fig:figureS3}.

The low-temperature limit allows us to get analytical results for both conventional and topological Josephson junctions, since we only have to determine the retarded and advanced Green's functions at zero energy by replacing $z = \pm i \eta$.
Keeping in mind that $\eta \to 0^+$ in the end, we also perform an expansion for small $\eta$.
To keep the analysis simple, we use $\Gamma_S := \Gamma_{T,L} = \Gamma_{T,R} = \Gamma_{B,L}= \Gamma_{B,R}$ and focus on the two cases $\Gamma_{\mathrm{CPS}} = 0$ and $\Gamma_{\mathrm{CPS}} = \Gamma_S$ for phases $\phi = \pi$ and $\phi \neq \pi$.

\begin{figure}
	\centering
	\includegraphics{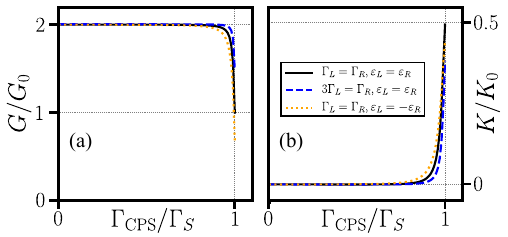}
	\caption{
		Robustness of the (a) electrical and (b) thermal conductance quantization in topological Josephson junctions for increasing CPS.
		Parameters: 
		$w = 3\Delta$,
		$\varepsilon_L = -\Delta/2$, 
		$\eta = 10^{-5}\Delta$,
		$k_B T = 10^{-3}\Delta$,
		$\Gamma_L = \Delta/5$,
		$\Gamma_S = \Delta$,
		$\phi = \pi$.
	}
	\label{fig:figureS4}
\end{figure}

For a topological Josephson junction, evaluating the transmission functions in the presence of a MZM requires special care, since the result changes drastically for $\phi = \pi$.
The transverse transmission functions  become
\begin{widetext}
\begin{subequations}
	\begin{align}
		\mathcal{T}_{RL}^{ee}(0) &= 
		\left\{\begin{array}{ccl}
			\mathcal{O}(\eta^4)
			& , & 
			\text{$\phi \neq \pi$ and $\Gamma_{\mathrm{CPS}} = 0$}
			\\
			\mathcal{O}(\eta^2) 
			& , & 
			\text{$\phi = \pi$ and $\Gamma_{\mathrm{CPS}} = 0$}
			\\
			\\
			\frac{4 \Gamma_L \Gamma_R }{(\Gamma_L + \Gamma_R)^2 + (\varepsilon_L + \varepsilon_R + 2 w)^2}
			+
			\mathcal{O}(\eta)
			& , & 
			\text{$\phi \neq \pi$ and $\Gamma_{\mathrm{CPS}} = \Gamma_S$}
			\\ \\
			\frac{
				\Gamma_L \Gamma_R \left( \Gamma_L^2+( \varepsilon_L + w)^2 \right) \left(\Gamma_R^2+(\varepsilon_R + w)^2 \right)
			}{
				\left[ 
				\Gamma_L \left(\Gamma_R^2+(\varepsilon_R + w)^2\right)
				+ \Gamma_R \left(\Gamma_L^2+(\varepsilon_L + w)^2\right) 
				\right]^2
			}
			+ \mathcal{O}(\eta)
			& , & 
			\text{$\phi = \pi$ and $\Gamma_{\mathrm{CPS}} = \Gamma_S$}
		\end{array} \right. ,
		\\
		\nonumber 
		\\
		\mathcal{T}_{RL}^{eh}(0) &= 
		\left\{\begin{array}{ccl}
			0
			& , & 
			\text{$\phi = 0$}
			\\
			\mathcal{O}(\eta^6)
			& , & 
			\text{$\phi \neq 0,\pi$ and $\Gamma_{\mathrm{CPS}} = 0$}
			\\
			\mathcal{O}(\eta^2) 
			& , & 
			\text{$\phi = \pi$ and $\Gamma_{\mathrm{CPS}} = 0$}
			\\
			\mathcal{O}(\eta^2)
			& , & 
			\text{$\phi \neq 0,\pi$ and $\Gamma_{\mathrm{CPS}} = \Gamma_S$}
			\\ \\ 
			\frac{\Gamma_L \Gamma_R \left(\Gamma_L^2+(\varepsilon_L+w)^2\right) \left(\Gamma_R^2+(\varepsilon_R+w)^2\right)}{
				\left[
				\Gamma_L \left(\Gamma_R^2+(\varepsilon_R+w)^2\right)
				+\Gamma_R \left(\Gamma_L^2+(\varepsilon_L+w)^2\right)
				\right]^2
			} + \mathcal{O}(\eta)
			& , & 
			\text{$\phi = \pi$ and $\Gamma_{\mathrm{CPS}} = \Gamma_S$}
		\end{array} \right. ,
		\\
		\nonumber 
		\\
		\mathcal{T}_{RR}^{eh}(0) &= 
		\left\{\begin{array}{ccl}
			0
			& , & 
			\text{$\phi = 0$}
			\\
			\mathcal{O}(\eta^2)
			& , & 
			\text{$\phi \neq 0,\pi$ and $\Gamma_{\mathrm{CPS}} = 0$}
			\\
			1 + \mathcal{O}(\eta) 
			& , & 
			\text{$\phi = \pi$ and $\Gamma_{\mathrm{CPS}} = 0$}
			\\
			\mathcal{O}(\eta^2) 
			& , & 
			\text{$\phi \neq 0,\pi$ and $\Gamma_{\mathrm{CPS}} = \Gamma_S$}
			\\
			\\
			\frac{
				\Gamma_R^2 \left(\Gamma_L^2 +(\varepsilon_L+w)^2 \right)^2
			}{
				\left[
				\Gamma_L \left(\Gamma_R^2 
				+ (\varepsilon_R+w)^2\right)
				+ \Gamma_R \left(\Gamma_L^2 + (\varepsilon_L+w)^2\right) 
				\right]^2
			} + \mathcal{O}(\eta)
			& , & 
			\text{$\phi = \pi$ and $\Gamma_{\mathrm{CPS}} = \Gamma_S$}
		\end{array} \right. ,
	\end{align}
\end{subequations}
which results in the transmissions [cf.~Eq.~\eqref{eq:equation4} in the main text]
\begin{subequations}
\begin{align}
	\mathcal{T}_{11}(0) &= 
	\left\{\begin{array}{ccl}
		\mathcal{O}(\eta^4)
		& , & 
		\text{$\phi = 0$ and $\Gamma_{\mathrm{CPS}} = 0$}
		\\ 
		2 + \mathcal{O}(\eta)
		& , & 
		\text{$\phi = \pi$ and $\Gamma_{\mathrm{CPS}} = 0$}
		\\ 
		\mathcal{O}(\eta^2)
		& , & 
		\text{$\phi \neq 0,\pi$ and $\Gamma_{\mathrm{CPS}} = 0$}
		\\ \\
		\frac{4 \Gamma_L \Gamma_R}{(\Gamma_L+\Gamma_R)^2+(\varepsilon_L+\varepsilon_R+2 w)^2} + \mathcal{O}(\eta)
		& , & 
		\text{$\phi \neq \pi$ and $\Gamma_{\mathrm{CPS}} = \Gamma_S$}
		\\ \\
		\frac{2 \Gamma_R \left(\Gamma_L^2+(\varepsilon_L+w)^2\right)}{
			\Gamma_L \left( \Gamma_R^2 + (\varepsilon_R+w)^2 \right)
			+\Gamma_R \left(\Gamma_L^2+(\varepsilon_L+w)^2\right)
		} + \mathcal{O}(\eta)
		& , & 
		\text{$\phi = \pi$ and $\Gamma_{\mathrm{CPS}} = \Gamma_S$}
	\end{array} \right. ,
	\\
	\nonumber 
	\\
	\mathcal{T}_{22}(0)  &= 
	\left\{\begin{array}{ccl}
		\mathcal{O}(\eta^4)
		& , & 
		\text{$\phi = 0$ and $\Gamma_{\mathrm{CPS}} = 0$}
		\\
		\mathcal{O}(\eta^2)
		& , & 
		\text{$\phi = \pi$ and $\Gamma_{\mathrm{CPS}} = 0$}
		\\ 
		\mathcal{O}(\eta^4)
		& , & 
		\text{$\phi \neq 0,\pi$ and $\Gamma_{\mathrm{CPS}} = 0$}
		\\  \\
		\frac{4 \Gamma_L \Gamma_R}{(\Gamma_L+\Gamma_R)^2+(\varepsilon_L+\varepsilon_R+2 w)^2} + \mathcal{O}(\eta)
		& , & 
		\text{$\phi \neq \pi$ and $\Gamma_{\mathrm{CPS}} = \Gamma_S$}
		\\ \\
		\frac{2 \Gamma_L \Gamma_R \left(\Gamma_L^2+(\varepsilon_L+w)^2\right) \left(\Gamma_R^2+(\varepsilon_R+w)^2\right)}{
			\left[
			\Gamma_L \left(\Gamma_R^2+(\varepsilon_R+w)^2\right)
			+\Gamma_R \left(\Gamma_L^2+(\varepsilon_L+w)^2\right)
			\right]^2
		} + \mathcal{O}(\eta)
		& , & 
		\text{$\phi = \pi$ and $\Gamma_{\mathrm{CPS}} = \Gamma_S$}
	\end{array} \right. .
\end{align}
\end{subequations}
\end{widetext}
We see that the electrical conductance is negligible at $\phi \neq \pi$ for $\Gamma_{\mathrm{CPS}} = 0$ and it is quantized with $G = 2G_0$ at $\phi = \pi$, similar to the result in Ref.~\cite{Yeyati_PRB_16} for a topological superconductor-normal metal junction.
Furthermore, we see that $G$ is completely determined by pure local Andreev reflection at the MZM. 
As presented in Fig.~\ref{fig:figureS4}(a), this conductance quantization is very robust for a wide range of $0 \leq \Gamma_{\mathrm{CPS}} \leq \Gamma_S$, with a suppression only for strong $\Gamma_{\mathrm{CPS}} \to \Gamma_S$.

For $\Gamma_{\mathrm{CPS}} = \Gamma_S$, the electrical conductance at $\phi \neq \pi$ follows the standard Lorentzian result for electron cotunneling through a resonant level at the energy $\varepsilon_L+\varepsilon_R+2 w$ \cite{cuevas2nd}.
Hence, it is completely dominated by electron cotunneling through the resonant level that is visible in Fig.~\ref{fig:figure2} in the main text. 
At $\phi = \pi$, the general electrical conductance is a mixture of all tunneling processes.
However, for the symmetric choice of parameters $\varepsilon_L = \varepsilon_R$ and $\Gamma_L = \Gamma_R$, the conductance is $G = G_0$, which is also shown in Fig.~\ref{fig:figureS4}(a).

On the other hand, as shown in Fig.~\ref{fig:figureS4}(b), the thermal conductance at $\phi = \pi$ is negligible at weak CPS.
This is due to the absence of LAR processes in the thermal conductance.
Similar to $G$ at $\Gamma_\mathrm{CPS} = \Gamma_S$, the precise value is not quantized and it is generally a mixture of EC and CAR contributions. 
However, for the symmetric choice of parameters $\varepsilon_L = \varepsilon_R$ and $\Gamma_L = \Gamma_R$, the thermal conductance is $K = K_0/2$, which is also shown in Fig.~\ref{fig:figureS4}(b).

In contrast to the topological case, the transverse transmission functions for a conventional Josephson junction are smooth functions of the phase difference $\phi$. 
We will not present the similar results for the $s$-wave case since the general results are (i) too cumbersome and (ii) not particularly insightful.
However, it is worth noting that, similar to the topological Josephson junction at $\phi = 0$, LAR and CAR are zero at $\phi = \pi$ for conventional Josephson junctions due to the symmetric choice of the couplings to the superconductors.

\bibliography{refs}

\end{document}